\documentclass[prb,twocolumn]{revtex4}
\usepackage[a4paper,top=0.75in,bottom=0.75in,left=0.75in,right=0.75in]{geometry}
\usepackage{amsmath}
\usepackage{amssymb}
\usepackage{mathptmx}
\usepackage[]{hyperref}
\usepackage[pdftex]{graphicx}
\usepackage[]{hyperref}
\def\nicefrac#1#2{\genfrac{}{}{}{1}{#1}{#2}}
\newcommand{\He}{$^3$He}
\newcommand{\Hea}{$^3$He-A}
\newcommand{\Hefour}{$^4$He}

\newcommand{\sro}{Sr$_{2}$RuO$_{4}$}
\def\cC{{\cal C}}

\newcommand{\grad}{\mbox{\boldmath$\nabla$}}
\newcommand{\curl}{\mbox{\boldmath$\nabla$}\times}
\newcommand{\be}{\begin{equation}}
\newcommand{\ee}{\end{equation}}
\newcommand{\ber}{\begin{eqnarray}}
\newcommand{\eer}{\end{eqnarray}}
\def\ket#1{\mbox{$\displaystyle\vert\,#1\,\rangle$}}

\begin{document}
\title{
\vspace*{-1.75cm}
\hspace*{-10cm}
{\bf\sl\footnotesize Published version: Viewpoint in Physics 9, 148 (2016) 
\href{http://dx.doi.org/10.1103/Physics.9.148}{[doi: 10.1103/Physics.9.148]}
}
\\
\vspace*{1.0cm}
Half-Quantum Vortices in Superfluid Helium
}
\author{J. A. Sauls}
\affiliation{Department of Physics and Astronomy, Northwestern University, Evanston, IL 60208 USA}
%\email{sauls@northwestern.edu}
\date{November 15, 2016}
\begin{abstract}
This is the first version of the \emph{Physics ViewPoint} \href{http://physics.aps.org/articles/v9/148}{[doi:10.1103/Physics.9.148]} article introducing the paper ``Observation of half-quantum vortices in topological superfluid \He'', by S. Autti et al., Low Temperature Laboratory, Department of Applied Physics, Aalto University, Finland, which appeared in \emph{Physical Review Letters}, December 14, 2016.\cite{aut16}
\end{abstract}
\maketitle
%-----------------------------------------------------------------------------------

%\noindent\underline{The Discovery of HQVs in a new phase of superfluid \He}

{\it Discoveries:} Researchers at the Low Temperature Laboratory in Finland succeeded in creating and manipulating vortices in a newly discovered superfluid phase of liquid \He, the light isotope of Helium. Remarkably the mass circulation of the superfluid velocity of each quantized vortex in this new state of matter is \emph{one half} the usual quantum of circulation predicted for vortices in a superfluid,\cite{ons49} i.e. $\nicefrac{1}{2}\left(h/2\,m_3\right)$, where $h$ is Planck's constant and $m_3$ is the mass of the \He\ atom. The discovery of half-quantum vortices (HQVs) in superfluid \He\ comes nearly 35 years after quantized vortices with unit circulation were first detected in rotating superfluid \He,\cite{hak82} and 40 years since half-quantum vortices were predicted theoretically as a new class of topological defects in condensed matter.\cite{vol76}
That prediction, combined with more recent theoretical ideas for developing topological condensed matter as platforms for quantum information processing, led to searches for HQVs in diverse condensed matter systems, from Bose-Einstein condensates of optically trapped spin S=1 $^{23}$Na atoms\cite{seo15} to spin-triplet superconductors thought to be electronic analogs of superfluid \Hea.\cite{jan11} 

%\noindent\underline{Background: Quantized Vortices \& Topological Defects}

%{\it Background:} 
Superfluids are quantum phases of matter in which the atoms condense into a single macroscopically occupied quantum state, i.e. the matter as a whole is described by a complex wave function, $\Psi=|\Psi|\,e^{i\vartheta}$, defined by an amplitude, $|\Psi|$, and phase $\vartheta$. As such the elementary constituents that condense are necessarily Bosons. In the case of Fermi liquids like \He\ the Bosons are bound Cooper pairs of spin $\nicefrac{1}{2}$ Fermions. 
Mass flow is described by the velocity field of the Cooper pairs, $\vec{v}=\nicefrac{\hbar}{2m_3}\grad\vartheta$, describing irrotational, potential flow, i.e. $\curl\vec{v} = 0$.
For a classical fluid confined in a vessel rotating with angular velocity $\vec\Omega$ the equilibrium state is co-rotation of the fluid with the walls of the vessel - a flow state with circulation $\curl\vec{v}=2\vec\Omega$.
Superfluids rotate by ``perforating the condensate'' with lines on which the wave function vanishes. These lines define singularities, or ``phase vortices'', characterized by the winding number, $N_{\cC}$, of the phase around any path $\cC$ that encloses the singular line, $\Delta\vartheta_{\cC}\equiv\oint_{\cC}\grad\vartheta\cdot d\vec{l} = N_{\cC}\,\times 2\pi$.
The physical requirement that the wave function for the condensate particles be a single-valued, constrains $N_{\cC}$ to take only integer values, which leads to quantization of the fluid circulation, $\oint_{\cC}\,\vec{v}\cdot d\vec{l}=N_{\cC}\,\frac{h}{2m_3}$, with elementary quantum of circulation given by $h/2m_3$.\cite{ons49} 

{\it HQVs:} In superfluid \He\  vortices with \emph{half} a quantum of circulation, i.e. $h/4m_3$, or a phase winding of $\oint_{\cC} \grad\vartheta \cdot d\vec{l} = \pi$ are possible.\cite{vol76} In the Bose liquid \Hefour, or Bose-Einstein condensate of spinless atoms, such vortices are prohibited by the requirement that $\Psi$ be single-valued. So, \emph{why} are HQVs allowed in superfluid \He, and \emph{how} were they revealed in the newly discovered Polar phase of \He?
The answer to the ``why'' is that there is a ``hidden half'' to the topology of the HQV - an extra $\pi$ twist - that is not related to circulation of the mass current, but rather to circulation of spin current.

Cooper pairs in superfluid \He\ are Bosons with orbital angular momentu $L=1$ and total nuclear spin $S=1$, the latter comprised of the two spin $s=\nicefrac{1}{2}$ \He\ nuclear spins. The newly discovered Polar phase of \He\ confined in Nafen\cite{dmi15} - a highly porous nematic aerogel - is a particular realization in which the orbital state is aligned with the axis $\hat{z}$ of the strands of the nematic aerogel, i.e. the p-state with orbital momentum ${p}_z$, and an equal amplitude superposition of Cooper pairs with opposite spin (ESP) projections,\cite{aoy06}
\be\label{eq-ESP}
\ket{\Psi} 
= 
|\Psi|\,e^{i\vartheta}\,
\left( 
e^{i\alpha}\ket{\downarrow\downarrow}
-
e^{-i\alpha}\ket{\uparrow\uparrow} 
\right) 
\,.
\ee
The phase angle, $\alpha$, between the two components of the spin state of the Cooper pairs is the ``hidden'' phase allowing spinor condensates to accomodate HQVs. The HQV with $\Delta\vartheta_{\cC}=\pi$ is realized by combining the phase change with a $\pi$ phase change of the internal spin state of the spinor condensate, $\Delta\alpha_{\cC}$, to ensure the total wave function is everywhere single-valued.
Alternatively, the HQV is a ``conventional'' $2\pi$ phase vortex in only one of the spin components, e.g.
$\Psi_{\downarrow\downarrow}=|\Psi|e^{i\vartheta_{\downarrow\downarrow}}$ and
$\Psi_{\uparrow\uparrow}=|\Psi|e^{i\vartheta_{\uparrow\uparrow}}$, with 
$\Delta\vartheta_{\downarrow\downarrow}=\Delta(\vartheta+\alpha)=2\pi$
and $\Delta\vartheta_{\uparrow\uparrow}=\Delta(\vartheta-\alpha)=0$.
This also highlights the facts that (i) in addition to the mass current circulating a HQV, there is a circulating spin current and (ii) that there are two HQVs with the same mass circulation, $\nicefrac{1}{2}(h/2m_3)$, but oppositely polarized spin currents.

%--------------------------------------------------------------------------------------------
\begin{figure}[t]
\includegraphics[width=0.8\columnwidth]{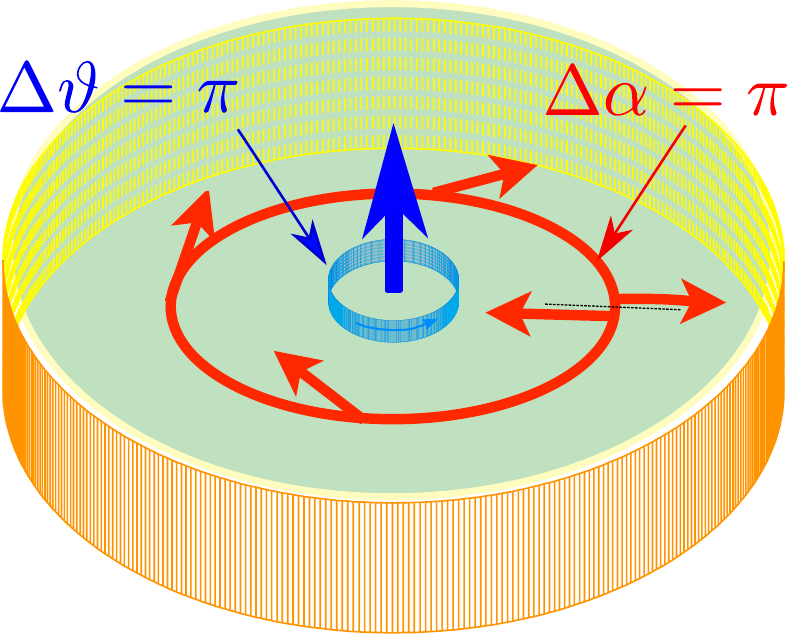}
\caption{In superfluid \He\ half-quantum vortices with circulation equal to $h/4m_3$
	correspond to a phase winding of $\Delta\vartheta_{\cC}=\pi$, combined with a rotation of the spin direction, 
        $\hat{d}$, by $\Delta\alpha_{\cC} = \pi$, giving a total phase winding of $2\pi$ for the $\ket{\downarrow\downarrow}$ 
        spin component and zero phase winding for the $\ket{\uparrow\uparrow}$ spin component.
	}\label{fig-HQV}
\end{figure}
%--------------------------------------------------------------------------------------------

An equivalent representation of the HQV is obtained by expressing Eq. \ref{eq-ESP} as a spinor with spin projection $\hat{d}\cdot\vec{S}=0$ along the direction $\hat{d} = \cos\alpha\,\hat{x} + \sin\alpha\,\hat{y}$, that lies in the plane perpendicular to the quantization axis for the $\ket{\uparrow\uparrow}$ spinor.
Thus, the ESP state represents a condensate of anti-ferromagentically spin-aligned Cooper pairs. The two inequivalent HQVs then correspond to composite topological defects in which the global phase changes by, $\Delta\vartheta_{\cC}=\pi$, combined with rotation of $\hat{d}$ by $\Delta\alpha_{\cC}=\pm\pi$ (c.f. Fig. \ref{fig-HQV}).\cite{vol76}

%\noindent\underline{Magnetic detection of HQVs in Superfluid \He}

{\it NMR Detection:} The ability to create the Polar phase of superfluid \He\ by infusing the liquid into Nafen\cite{dmi15} was key to generating and stabilizing HQVs in superfluid \He.
Autti and colleauges created vortices by rotating superfluid \He\ confined in Nafen. They did not directly measure the mass circulation around each vortex to check if the vortices had half-quantum circulation. Rather, the key to the detection of HQVs by the Finnish team, as well as their stability, is the nuclear magnetism of superfluid \He.
The interaction between the nuclear magnetic moments of the \He\ atoms that comprise Cooper pairs is sensitive to the \emph{relative} orientation of spin quantization axis and the orbital axis of the molecular pairs, a fact that lead to the original identification of the superfluid phases of pure \He,\cite{leg73a} and to the development of NMR as a powerful spectroscopy of the spin and orbital wave function for the condensate of Cooper pairs under wide ranging conditions, and particularly for indentifying new superfluid phases in confined geometries and infused into nano-structures.\cite{pol12,dmi15}
For the Polar phase of \He\ the nuclear dipolar energy is neutralized in the interaction between HQVs with oppositely oriented spin currents, thus allowing for an array of HQVs to be established as the equilibrium state of the Polar phase of \He\ under rotation.\cite{min14}

The Finnish group utilized the spin structure of pairs of HQVs with opposite-spin currents to detect arrays of HQVs, by first creating HQVs under equilibrium conditions with the magnetic field aligned along the rotation axis and the nematic axis of the Polar phase. Once the array of HQVs were created the experimenters rotated the NMR field perpendicular to the nematic axis, generating 
a solitary wave, a ``soliton'', in the orientation of the spin axis connecting pairs of HQVs that is detected as a resonance peak shifted to a frequency below the Larmor frequency.\cite{min16}
This is the unique fingerprint of the pair of HQVs.

The researchers also relied on another key property of \He\ confined in nematic aerogel. Nafen is a random solid of rod-like impurites that ``pin'' the HQVs, preventing pairs of HQVs from combining into SQVs under rotation of the NMR field.
Thus, the interaction between HQVs and the anisotropic random field of the Nafen provides both stability for the newly discovered Polar phase of \He\ as well as the stability of the HQVs from recombination into magnetically inert single-quantum mass vortices.

%\noindent\underline{Looking Forward}

{\it Looking Forward:} The discovery of HQVs in superfluid \He\ is part of a larger trend using topology in condensed matter systems as an organizing principle for understanding quantum states of matter. 
Recent developments in nano-fabrication and nano-structured materials, combined with low temperature research on quantum fluids, have led to prediction and discovery of new topological phases of matter, and have opened up new research directions in confined and engineered quantum matter, including the discovery of the topological structure of HQVs in a newly discovered phase of superfluid \He, both of which previously existed only on paper and in the minds of few theoretical physicists. 
Indeed the existence, stability and means to create HQVs in superfluid \He, combined with experimental tools to manipulate these topological excitations, will likely open new directions for the study of related classes of topological structures in quantum fluids, and possibly stimulate search and discovery in a broad range of condensed matter systems.

{\it Acknowledgement:} The research of the author on topological properties of quantum matter is supported by the National Science Foundation, and currently by Grant DMR-1508730.

%---------------------------------------------------------------------------------
%\bibliographystyle{apsrev4-1_PRX_style}
%\bibliography{QFS,CM,Books}
%---------------------------------------------------------------------------------
%
\end{document}